\documentclass[preprint,12pt]{elsarticle}%
\usepackage{amssymb}
\usepackage{amsfonts}
\usepackage{amsmath}
\usepackage{graphicx}%
\providecommand{\U}[1]{\protect\rule{.1in}{.1in}}
\journal{journal}

\begin{document}
\bigskip\bigskip%
\begin{frontmatter}%


%

\title
{An approximate JKR solution for a general contact, including rough contacts}%

%

\author{M. Ciavarella}%
%

\address
{Politecnico di BARI. Center of Excellence in Computational Mechanics. Viale Gentile 182, 70126 Bari. Mciava@poliba.it}%
%

\begin{abstract}%

In the present note, we suggest a simple closed form approximate solution to
the adhesive contact problem under the so-called JKR regime. The derivation is
based on generalizing the original JKR energetic derivation assuming
calculation of the strain energy in adhesiveless contact, and unloading at
constant contact area. The underlying assumption is that the contact area
distributions are the same as under adhesiveless conditions (for an
appropriately increased normal load), so that in general the stress intensity
factors will not be exactly equal at all contact edges. The solution is simply
that the indentation is $\delta=\delta_{1}-\sqrt{2wA^{\prime}/P^{\prime\prime
}}$ where $w$ is surface energy, $\delta_{1}$ is the adhesiveless indentation,
$A^{\prime}$ is the first derivative of contact area and $P^{\prime\prime}$
the second derivative of the load with respect to $\delta_{1}$. The solution
only requires macroscopic quantities, and not very elaborate local
distributions, and is exact in many configurations like axisymmetric contacts,
but also sinusoidal waves contact and correctly predicts some features of an
ideal asperity model used as a test case and not as a real description of a
rough contact problem. The solution permits therefore an estimate of the full
solution for elastic rough solids with Gaussian multiple scales of roughness,
which so far was lacking, using known adhesiveless simple results. The result
turns out to depend only on rms amplitude and slopes of the surface, and as in
the fractal limit, slopes would grow without limit, tends to the adhesiveless
result -- although in this limit the JKR model is inappropriate. The solution
would also go to adhesiveless result for large rms amplitude of roughness
$h_{rms}$, irrespective of the small scale details, and in agreement with
common sense, well known experiments and previous models by the author.%

\end{abstract}%
%

%

\end{frontmatter}%



\section{\bigskip Introduction}

Exact solution to adhesive problems are very scarse. Bradley (1932) and
Derjaguin (1934) obtained the adhesive force between two \textit{rigid}
spheres, equal to $2\pi Rw$, where $w$ is the work of adhesion, and $R$ is the
radius of the sphere. Then, JKR (Johnson Kendall and Roberts 1971) developed
the first exact theory for elastic bodies, namely spheres, assuming adhesive
forces occur entirely within the contact area, obtaining $3/4$ of the Bradley
pull-off value, and independence on the elastic modulus which seem to indicate
that the result would be corresponding to the rigid Bradley limit. The result
was even more surprising when Derjaguin-Muller-Toporov (DMT) developed their
elastic theory (Derjaguin \textit{et al.}, 1975) which seemed to indicate the
same pull-off value of Bradley rather that JKR. Tabor brilliantly solved the
dilemma, indicating transition from rigid to JKR depends on the Tabor
parameter (Tabor, 1977)\bigskip%
\begin{equation}
\mu=\left(  \frac{Rw^{2}}{E^{\ast2}a_{0}^{3}}\right)  ^{1/3} \label{Tabor}%
\end{equation}
where $a_{0}$ is the range of attraction of adhesive forces, close to atomic
distance for crystals, and $E^{\ast}$ the plane strain elastic modulus.

The JKR regime therefore is only valid for large Tabor parameters (especially
if instability at jump-into contact is required accurately, see Ciavarella
\textit{et al.}2017). JKR permits to find many solutions easily by
superposition of contact and crack solutions (see Johnson, 1995), whereas the
original JKR energetic method has been less popular except of course the
adhesion problem can be formulated in elaborate numerical algorithms by
minimization methods (Carbone \textit{et al.}, 2015). Particularly the problem
of rough surfaces has seen significant effort in the last 40 years or so
(Fuller \& Tabor, 1975, Persson, 2002, Pastewka \&\ Robbins, 2014, Persson and
Scaraggi, 2014, Ciavarella, 2015, Afferrante \textit{et al.}, 2015, Ciavarella
and Papangelo, 2017a,b,c, Ciavarella \textit{et al.}, 2017, 2018, Ciavarella,
2017a,b), but no simple theories exist which permit to estimate the JKR
regime, including negative loads and pull-off, except for Fuller \& Tabor
(1975) asperity theory, which however has been questioned by Pastewka
\&\ Robbins (2014), and certainly contains many strong approximations inherent
in the asperity model. In the DMT regime, a very simple solution was given by
Ciavarella (2017a) with a "bearing-area" model, which turned out to give very
reasonable fit of the Pastewka \&\ Robbins (2014) pull-off data, whereas some
discrepancy was remarked about the area-slope "stickiness" criterion with
their own pull-off data. Persson (2002) is aimed at the JKR regime, seems
perfectly reversible, is quite complex and anyway it is probably not valid in
unloading as shown in the plots in Persson and Scaraggi (2014) which only show
the positive load regime -- and also as explained in details by Carbone et al.
(2015) who have constructed for 1D profiles, loading and unloading curves and
PSD (Power Spectrum Density) of the deformed profile closely follows a power
law predicted by Persson's theory, but not on unloading. In particular, they
explain why Persson's theory is not adequate for adhesion. Moving to the DMT
approximation of Persson and Scaraggi (2014), the contact is assumed to be
split into "repulsive" contact areas and "attractive" contact areas, and no
effect of tensile tractions occurs so there is a simple convolution of
separation of the repulsive solution with the force-separation law  ---
however, the DMT approximation leads to large errors even in the simple case
of a sphere or a cylinder (Ciavarella, 2017b), and it is unclear what happens
for rough contacts where many further approximations are made. In any case,
the solution remains numerical and not simple in this case either.

All these models are purely theoretical or numerical. Experimental studies
typically rely on spherical geometry like Fuller \& Tabor (1975). 

JKR (1971) originally derived an energetic method which could serve as an
approximate solution to a much more general contact case, not just including
halfspace geometries but really anything for which we know the adhesionless
solution. We shall therefore generalize the JKR model to arbitrary contact
geometry, in an approximate sense, in the present paper.

\section{The model}

We need to consider the total potential energy of the system comprising
elastic strain energy, surface energy, and [when load $P$ is prescribed]
potential energy of the applied force.

The elastic strain energy can be determined by devising the original JKR
loading scenario leading to the required final state and calculating the work
done during loading. Such scenario is suggested by the superposition in
comprising the two steps (see Fig.1)

\begin{itemize}
\item (i) "repulsive" loading without adhesive forces until the contact area
is a given value (this is the load path $\overline{OA}$ as in the original JKR
paper), followed by

\item (ii) rigid-body displacement at constant total contact area $A$ (this is
load path $\overline{AB}$ as in the original JKR paper) of an unknown amount
which we shall find by a minimization procedure of the total potential, like
in the classical Griffith crack problem. In Irwin's equivalent procedure, this
unloading could be prescribed until the required stress-intensity factor (SIF)
is achieved at the contact edge. Here, we can only fulfill this requirement in
an "average" sense, because the SIF at the indivual contact edges will differ.
But it is not convenient to evaluate the individual SIFs and make their
average, as this would require in general a cumbersome procedure. Similarly, a
precise solution should minimize over many variables, which are the position
and size and number of the contact spots (Carbone \textit{et al.}, 2015)
\end{itemize}

\begin{center}%
\begin{tabular}
[c]{ll}%
{\includegraphics[
height=3.3728in,
width=5.0571in
]%
{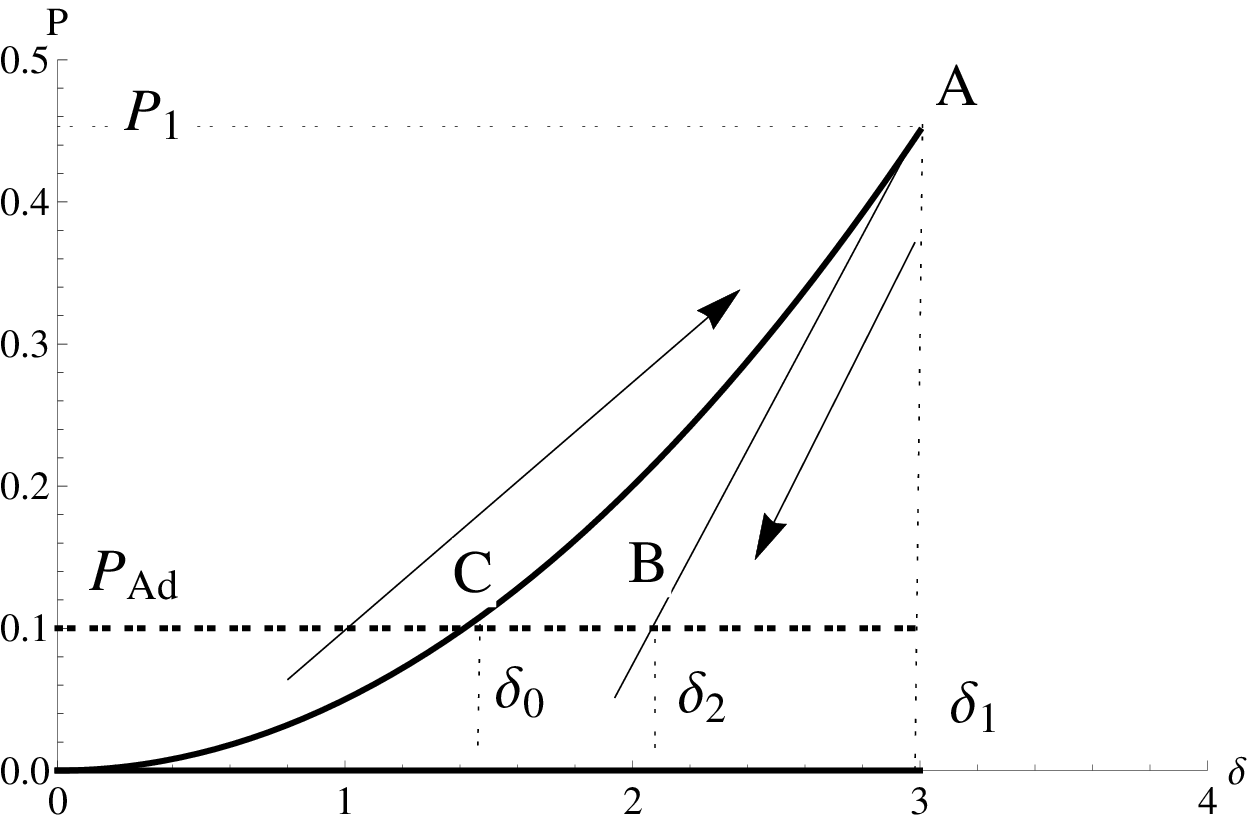}%
}
&
\end{tabular}

Fig.1. The loading scenario. (i) "repulsive" loading without adhesive forces
until the contact area is a given value (load path $\overline{OA}$ as in the
original JKR paper); (ii) rigid-body displacement at constant total contact
area $A$ (load path $\overline{AB}$ as in the original JKR paper)
\end{center}

In the present paper, we repeat this process in general, and not just for the
spherical Hertzian contact as in the original JKR paper. During phase (i), we
can integrate the load-displacement curve to get the elastic strain energy%
\begin{equation}
U_{1}\left(  \delta_{1}\right)  =\int_{0}^{\delta_{1}}P\left(  \delta\right)
d\delta
\end{equation}
while in the second phase (ii) we have an unknown downloading to
$\delta=\delta_{2}$
\begin{equation}
U_{2}\left(  \delta_{1},\delta\right)  =\int_{\delta_{1}}^{\delta}P_{A}\left(
\delta\right)  d\delta
\end{equation}
where we write $P_{A}\left(  \delta\right)  $ for unloading as we keep the
contact area $A$ fixed.

We then obtain the total elastic strain energy is
\begin{equation}
U\left(  \delta_{1},\delta\right)  =U_{1}\left(  \delta_{1}\right)
-U_{2}\left(  \delta_{1},\delta\right)
\end{equation}
However, since the contact area is fixed in the second phase, the load really
decreases linearly so the Taylor series expansion can be exactly truncated to
first order
\begin{equation}
P_{Ad}\left(  \delta\right)  =P_{1}+\left(  \frac{\partial P}{\partial\delta
}\right)  _{\delta_{1}}\left(  \delta-\delta_{1}\right)  \label{adhesive-load}%
\end{equation}
and hence the work done is%
\begin{equation}
U_{2}\left(  \delta_{1},\delta\right)  =\int_{\delta_{1}}^{\delta}%
P_{Ad}\left(  \delta\right)  d\delta=P_{1}\left(  \delta-\delta_{1}\right)
+\left(  \frac{\partial P}{\partial\delta}\right)  _{\delta_{1}}\frac{\left(
\delta-\delta_{1}\right)  ^{2}}{2}%
\end{equation}

\subsection{Displacement control}

The surface energy is $-Aw$, and hence if the displacement is prescribed, the
total potential energy is%
\begin{equation}
\Pi\left(  \delta_{1},\delta\right)  =U\left(  \delta_{1},\delta\right)  -Aw
\end{equation}

In practice, we would need to write $\Pi$ as a function of the contact area
$A$ since the equilibrium position is then determined by the condition%
\begin{equation}
\frac{\partial\Pi}{\partial A}=0
\end{equation}

However, this can be rewritten obviously using the chain's rule, as%
\begin{equation}
\frac{\partial U}{\partial\delta_{1}}\frac{\partial\delta_{1}}{\partial A}=w
\end{equation}

Now, trivially we find
\begin{align*}
\frac{\partial U_{1}}{\partial\delta_{1}}  &  =P_{1}\\
\frac{\partial U_{2}}{\partial\delta_{1}}  &  =-P_{1}+\frac{\partial P_{1}%
}{\partial\delta_{1}}\left(  \delta-\delta_{1}\right)  +\frac{\partial\left[
\left(  \frac{\partial P}{\partial\delta}\right)  _{\delta_{1}}\right]
}{\partial\delta_{1}}\frac{\left(  \delta-\delta_{1}\right)  ^{2}}{2}-\left(
\frac{\partial P}{\partial\delta}\right)  _{\delta_{1}}\left(  \delta
-\delta_{1}\right)
\end{align*}

Summing the contribution and cancelling the equal terms, this reduces to%
\[
\frac{\partial^{2}P}{\partial\delta_{1}^{2}}\frac{\left(  \delta-\delta
_{1}\right)  ^{2}}{2}-w\frac{\partial A}{\partial\delta_{1}}=0
\]
which solves into
\begin{equation}
\delta=\delta_{1}-\sqrt{2w\frac{\partial A}{\partial\delta_{1}}/\frac
{\partial^{2}P}{\partial\delta_{1}^{2}}} \label{general}%
\end{equation}
as we have choosen only the physical solution for $\delta<\delta_{1}$. This
(\ref{general}) is the quite general approximate solution, valid \textit{for
any arbitrary contact problem}. Notice that it only requires macroscopic
quantities, and not very elaborate local distributions.

A solution under force control will not differ, but the stability condition
(which involves the second derivative of the potential) will be different.
However, we don't need to describe this trivial extension.

\subsection{Check for Hertzian contact, and other validations}

For Hertzian contact,
\[
\frac{\pi a^{2}}{\pi R}=\frac{A}{\pi R}=\delta_{1},\quad P=\frac{4}{3}E^{\ast
}\sqrt{R}\delta_{1}^{3/2},\quad
\]
and hence from our general result (\ref{general})
\begin{equation}
\delta=\delta_{1}-\sqrt{\frac{2w\pi\sqrt{R}}{E^{\ast}}\delta_{1}^{1/2}}
\label{result}%
\end{equation}
which reduces to the known form of the exact JKR solution (Johnson et al,
1971), when translated back into an expression for indentation vs contact
radius%
\[
\delta=\frac{a^{2}}{R}-\sqrt{\frac{2w\pi}{E^{\ast}}a}%
\]

This exact coincidence of the proposed result (\ref{general}) with the JKR
solution would also occur for any axisymmetric contact case (see Popov and
He\ss , 2015), since obviously the contact area remains circular, including
the case of a waviness when contact remains compact, like in Guduru (2007),
which shows a possible large enhancement of adhesion with respect to the
smooth sphere case. As in this case there is no approximation in our
calculation with respect to any axisymmetric configuration, the SIF will be
constant at circular contact edge by construction, and there is no need to
further test these cases.

Another case where the solution would be exact is the sinusoidal contact of
Johnson (1995), since here the contact area is by simmetry defined by a single
parameter, and a fortiori the SIF will be equal for the configurations
described by this single parameter. Hence, there is no check the result in
this case either. For 2D sinusoidal contact, instead, even of equal
wavelengths, an error may appear when contact area is large, since the
contacts will not be circular. A comparison with an asperity model follows.

\subsection{Comparison - asperity models}

Real surfaces do not satisfy the approximation of asperity models, because of
geometrical errors in the description of the roughness, and also because of
neglecting interaction effects. However, in an ideal case of a true set of
independent asperities, if our result were exactly valid, one should be able
to obtain the same result by the classical approach of superposing results for
each individual asperity, and our direct equation. Therefore, this comparison
is a valid test case to check the effect of a distribution of contact spots of
different sizes.

We take for simplicity the exponential distribution of heights $\phi=\frac
{1}{\sigma_{s}}\exp\left(  -\frac{z_{s}}{\sigma_{s}}\right)  (z_{s}>0)$, which
is good enough for our test. Area and load are each proportional to number of
asperities in contact $n=N\exp\left(  -\frac{u}{\sigma_{s}}\right)  $, where
$N$ is their total number. Here, $\sigma_{s}$ is a scale parameter of the
heights. Hence, the ratio of area to load is constant in the classical
repulsive case%

\begin{align}
\qquad A &  =\pi R\sigma_{s}n=\pi R\sigma_{s}N\exp\left(  -\frac{u_{1}}%
{\sigma_{s}}\right)  \\
P &  =nE^{\ast}\left(  \sigma_{s}^{3}R\right)  ^{1/2}\sqrt{\pi}=E^{\ast
}\left(  \sigma_{s}^{3}R\right)  ^{1/2}\sqrt{\pi}N\exp\left(  -\frac{u_{1}%
}{\sigma_{s}}\right)
\end{align}
where $R$ is radius of the asperity. Hence,
\begin{align}
\frac{\partial A}{\partial\delta_{1}}  & =-\frac{\partial A}{\partial u}=\pi
RN\exp\left(  -\frac{u_{1}}{\sigma_{s}}\right)  \\
\frac{\partial^{2}P}{\partial\delta_{1}^{2}}  & =\frac{\partial^{2}P}{\partial
u^{2}}=E^{\ast}\frac{R^{1/2}}{\sigma_{s}^{1/2}}\sqrt{\pi}N\exp\left(
-\frac{u_{1}}{\sigma_{s}}\right)
\end{align}
and our equation (\ref{general}) gives%
\begin{equation}
\delta=\delta_{1}-\sqrt{2w\frac{\pi R^{1/2}\sigma_{s}^{1/2}}{E^{\ast}\sqrt
{\pi}\ }}%
\end{equation}

Now, the indentation and mean separation are in the relationship
\begin{equation}
\delta-\delta_{1}=u_{1}-u\label{indentation-separation}%
\end{equation}
and hence we have obtained the repulsive load vs actual adhesive gap. 

Now, we return to (\ref{adhesive-load}) which here will be
\begin{equation}
P_{Ad}=P_{1}+\left(  -\frac{\partial P}{\partial u_{1}}\right)  \left(
u_{1}-u\right)
\end{equation}
giving
\begin{equation}
P_{Ad}=P_{1}\left(  1-\sqrt{\theta_{\exp}}\right)
\end{equation}
where $\theta_{\exp}$ is the same as that predicted by Fuller Tabor in the
exponential form if we take DMT instead of JKR for the individual asperity
(Ciavarella \& Papangelo 2017d),
\begin{equation}
\theta_{\exp}=2\sqrt{\pi}\frac{R^{1/2}}{\sigma_{s}^{3/2}}l_{a}%
\end{equation}
and hence the contact is either always tensile or compressive, a result which
is exactly as obtained in asperity model of Fuller and Tabor (1975) in the
exponential form (Ciavarella \& Papangelo 2017d). However, the asperity model
would give $P_{Ad}=P_{1}\left(  1-\theta_{\exp}\right)  $ and hence the
adhesive load is different because of the square root. Notice however that we
have obtained a result which is non-hysteretic but is close to the DMT
asperity model which therefore is closer to the unloading regime of a JKR
solution than to a loading one.\ This gives us some confidence that the
results for a more general problem will be of some validity. In particular,
the transition from sticky to non-sticky seems correctly predicted but the
quantitative results seems to be that for low adhesion, the effect of adhesion
may be overestimated, while for high adhesion, underestimated. But some level
of approximation is inevitable in all models of complex adhesive problems, and
particularly in the case of roughness, where no exact solution is known even
just in the case without adhesion.

\section{Application to rough contacts - Persson's theory}

\bigskip In applying the model to a random rough surface, we need a theory to
estimate the variation of contact area, and of load, with indentation: the
terms $\frac{\partial A}{\partial\delta_{1}},\frac{\partial^{2}P}%
{\partial\delta_{1}^{2}}$ in (\ref{general}). Persson (2007) gives a mean
repulsive pressure $\sigma_{rep}$ vs mean separation $u$ law which, for the
practical case of self-affine surfaces of low fractal dimension ($D\simeq2.2$
is a value of common experience, Persson \textit{et al.}, 2014), assumes a
simple asymptotic form which is sufficiently valid for not too large
$\sigma_{rep}$
\begin{equation}
\frac{\sigma_{rep}}{E^{\ast}}\simeq\frac{3}{8\gamma}q_{0}h_{rms}\exp\left(
\frac{-u}{\gamma h_{rms}}\right)  \label{Persson1}%
\end{equation}
where $\gamma\simeq0.45$, $q_{0}$ is the smallest wavevector in the
self-affine process where the power law Power Spectrum starts, and $h_{rms}$
is the rms amplitude of roughness. Notice that we have corrected the
multiplier in agreement with numerical findings of Papangelo \textit{et
al.}(2017). Hence, for low fractal dimensions, the result does not depend on
fine-scale details of the surface.

Persson (2001) then suggests for the proportion of actual contact at a given
nominal pressure which, after a more recent corrective factor of Putignano et
al (2012) has been included, reads%

\begin{equation}
\frac{A_{rep}}{A_{0}}=\operatorname{erf}(\frac{\sqrt{\pi}}{2}\frac
{\sigma_{rep}}{\sigma_{rough}}) \label{Persson2}%
\end{equation}
where $\sigma_{rough}=E^{\ast}h_{rms}^{\prime}/2$ where $h_{rms}^{\prime}$ is
the rms slope of the surface, and $\sigma_{rep}$ can be estimated as a
function of $u$ from (\ref{Persson1}).

We can combine the two Persson's results (\ref{Persson1},\ref{Persson2}) to
find%
\begin{equation}
\frac{A_{rep}}{A_{0}}=\operatorname{erf}\left[  \frac{\sqrt{\pi}}{2}%
\frac{E^{\ast}}{\sigma_{rough}}\frac{3}{8\gamma}q_{0}h_{rms}\exp\left(
\frac{-u}{\gamma h_{rms}}\right)  \right]
\end{equation}
from which, putting
\begin{equation}
\alpha=\frac{\sqrt{\pi}}{2}\frac{E^{\ast}}{\sigma_{rough}}\frac{3}{8\gamma
}q_{0}h_{rms}%
\end{equation}
we obtain
\begin{equation}
\frac{\partial}{\partial u}\frac{A_{rep}}{A_{0}}=-\frac{2\alpha}{\gamma
h_{rms}\sqrt{\pi}}\exp\left(  -\alpha^{2}\exp\left(  -\frac{2u}{\gamma
h_{rms}}\right)  -\frac{u}{\gamma h_{rms}}\right)
\end{equation}
whereas
\begin{equation}
\frac{\partial}{\partial u}\frac{\sigma_{rep}}{E^{\ast}}\simeq-\frac
{3}{8\gamma^{2}}q_{0}\exp\left(  \frac{-u}{\gamma h_{rms}}\right)  \quad
,\quad\frac{\partial^{2}}{\partial u^{2}}\frac{\sigma_{rep}}{E^{\ast}}%
\simeq\frac{3}{8\gamma^{3}}q_{0}\frac{1}{h_{rms}}\exp\left(  \frac{-u}{\gamma
h_{rms}}\right)
\end{equation}
Finally, noticing that the indentation and mean separation are in the
relationship (\ref{indentation-separation}), we need to change sign to the
derivatives, and then we can substitute in the general solution.

Let us introduce the length $l_{a}=w/E^{\ast}$ as an alternative measure of
adhesion, and we can use again (\ref{Persson1}), such that $\left(  \log
\frac{8\gamma^{2}\sigma_{rep}}{3E^{\ast}q_{0}}\right)  \simeq-u_{1}$ to show
after some algebra that our general result (\ref{result}) leads to%
\begin{align}
u  &  =u_{1}+\sqrt{\frac{4\gamma}{\sqrt{\pi}}\frac{l_{a}h_{rms}}%
{h_{rms}^{\prime}}\exp\left(  -\frac{3\pi}{8\gamma}\frac{q_{0}h_{rms}}%
{h_{rms}^{\prime2}}\frac{\sigma_{rep}}{E^{\ast}}\right)  }\nonumber\\
&  =-\log\frac{8\gamma^{2}\sigma_{rep}}{3E^{\ast}q_{0}}+\sqrt{\frac{4\gamma
}{\sqrt{\pi}}\frac{l_{a}h_{rms}}{h_{rms}^{\prime}}\exp\left(  -\frac{3\pi
}{8\gamma}\frac{q_{0}h_{rms}}{h_{rms}^{\prime2}}\frac{\sigma_{rep}}{E^{\ast}%
}\right)  } \label{u}%
\end{align}
This solution is now written in terms of actual adhesive mean separation, as a
function of the repulsive pressure.

It is clear that in the fractal limit, this solution tends to the adhesiveless
result, because $h_{rms}^{\prime}\rightarrow\infty$. But it would also go to
adhesiveless for large $h_{rms}$ because of the term in the exponential.

To find pull-off, we need to elaborate more on the solution. We have to return
to the definition of the load after unloading at constant contact area, which
now for rough surfaces will read
\begin{equation}
\frac{\sigma_{Ad}}{E^{\ast}}=\frac{\sigma_{rep}\left(  u_{1}\right)  }%
{E^{\ast}}+\left(  \frac{1}{E^{\ast}}\frac{\partial\sigma_{rep}}{\partial
u}\right)  _{u_{1}}\left(  u-u_{1}\right)
\end{equation}
Also, from (\ref{Persson1}) $\left(  \frac{1}{E^{\ast}}\frac{\partial
\sigma_{rep}}{\partial u}\right)  _{u_{1}}=-\frac{3}{8\gamma^{2}}q_{0}%
\exp\left(  \frac{-u_{1}}{\gamma h_{rms}}\right)  $ and hence
\begin{align}
\frac{\sigma_{Ad}}{E^{\ast}} &  =\frac{\sigma_{rep}\left(  u_{1}\right)
}{E^{\ast}}-\frac{3}{8\gamma^{2}}q_{0}\exp\left(  \frac{-u_{1}}{\gamma
h_{rms}}\right)  \left(  u-u_{1}\right)  \\
&  =\frac{3}{8\gamma}q_{0}h_{rms}\exp\left(  \frac{-u_{1}}{\gamma h_{rms}%
}\right)  \left[  1-\frac{1}{\gamma h_{rms}}\left(  u-u_{1}\right)  \right]
\end{align}
and finally using (\ref{u}) for $\left(  u-u_{1}\right)  $
\begin{align}
\frac{\sigma_{Ad}\left(  u_{1}\right)  }{E^{\ast}} &  =\frac{3}{8\gamma}%
q_{0}h_{rms}\exp\left(  \frac{-u_{1}}{\gamma h_{rms}}\right)  \times
\nonumber\\
&  \left[  1-\frac{1}{\gamma h_{rms}}\sqrt{\frac{4\gamma}{\sqrt{\pi}}%
\frac{l_{a}h_{rms}}{h_{rms}^{\prime}}\exp\left(  -\left(  \frac{3}{8\gamma
}\right)  ^{2}\pi\frac{\left(  q_{0}h_{rms}\right)  ^{2}}{h_{rms}^{\prime2}%
}\exp\left(  \frac{-u_{1}}{\gamma h_{rms}}\right)  \right)  }\right]
\label{JKR-final}%
\end{align}
which gives the mean pressure with adhesion as a function of the "adhesionless
separation" $u_{1}$ which is only the value needed to obtain a given contact
area: but spanning all values of $u_{1}$, we can find anyway the entire
solution, as we can use (\ref{u}) to find the actual $u$, and all the other
quantities are known. We can also rewrite the solution introducing a constant
$a_{0}$ of the order of atomic spacing to obtain a non dimensional version.

In particular, but we shall see more problematic in terms of accuracy, is to
push the solution to find the minimum value of the adhesive stress, and hence
pull-off%
\begin{equation}
\left(  \sigma_{Ad}\right)  _{po}=\min\sigma_{Ad}\left(  u_{1}\right)
\end{equation}
not because the minimum requires a numerical root finder, but because it turns
out in practise that this approximation leads to either always tensile forces,
or always compressive. This is due to the fact that we have used the simplest
approximation of the Persson's solution, at small pressures (\ref{Persson1}),
whereas the full solution would require a much more elaborate form, which we
leave for further studies, when $u_{1}\rightarrow0$, and the repulsive
pressure tends to very high values. In fact the only possible simple estimate
for pull-off is at zero gap%
\begin{align}
\frac{\sigma_{Ad}\left(  0\right)  }{E^{\ast}} &  =\frac{3}{8\gamma}a_{0}%
q_{0}\frac{h_{rms}}{a_{0}}\times\nonumber\\
&  \left[  1-\frac{1}{\gamma h_{rms}/a_{0}}\sqrt{\frac{4\gamma}{\sqrt{\pi}%
}\frac{\left(  l_{a}/a_{0}\right)  \left(  h_{rms}/a_{0}\right)  }%
{h_{rms}^{\prime}}\exp\left(  -\left(  \frac{3}{8\gamma}\right)  ^{2}\pi
\frac{\left(  a_{0}q_{0}\frac{h_{rms}}{a_{0}}\right)  ^{2}}{h_{rms}^{\prime2}%
}\right)  }\right]
\end{align}
and obviously this is tensile only if
\begin{equation}
\gamma h_{rms}/a_{0}<\frac{4}{\sqrt{\pi}}\frac{\left(  l_{a}/a_{0}\right)
}{h_{rms}^{\prime}}\exp\left(  -\left(  \frac{3}{8\gamma}\right)  ^{2}\pi
\frac{\left(  a_{0}q_{0}\frac{h_{rms}}{a_{0}}\right)  ^{2}}{h_{rms}^{\prime2}%
}\right)
\end{equation}
which is a quite restrictive (implicit) condition on the rms amplitude, which
we can approximate for not too small $h_{rms}^{\prime}$ (notice that the
numerator in the exponential term is an apparent "slope" at small wavevectors
and hence is much smaller than the denominator)
\begin{equation}
h_{rms}/a_{0}<\left(  h_{rms}/a_{0}\right)  _{th}=\frac{4}{\sqrt{\pi}\gamma
}\frac{\left(  l_{a}/a_{0}\right)  }{h_{rms}^{\prime}}\label{new-criterion}%
\end{equation}

\subsection{Comparison with Pastewka-Robbins (PR) criterion}

PR stickiness criterion is obtained in the original paper (eqt.10), in the
form
\begin{equation}
\frac{h_{rms}^{\prime}\Delta r}{\kappa_{rep}l_{a}}\left[  \frac{h_{rms}%
^{\prime}d_{rep}}{4\Delta r}\right]  ^{2/3}<\pi\left(  \frac{3}{16}\right)
^{2/3}\simeq1.03\label{PReqt10}%
\end{equation}
where $\Delta r$ is range of attractive forces, and $d_{rep}$ is a
characteristic diameter of repulsive contact areas, which they estimate as
$d_{rep}=4h_{rms}^{\prime}/h_{rms}^{\prime\prime}\ $\ and finally
$\kappa_{rep}\approx2$. In order to incorporate their choice of truncated
potentials, the range of attraction is easily obtained from Suppl.Inf. of PR
paper to be $\Delta r/a_{0}=\sqrt{24l_{a}/a_{0}}.$ For the Lennard-Jones
situations ($l_{a}/a_{0}=0.05$), $\Delta r\approx a_{0}$ and grouping the
variables using the Nayak bandwidth parameter $\alpha_{N}=\frac{m_{0}m_{4}%
}{m_{2}^{2}}$, where $m_{n}$ are the moments of order $n$ in the random
process, we can restate\ (\ref{PReqt10}) as
\begin{equation}
\frac{h_{rms}}{a_{0}}<\sqrt{\alpha_{N}}\left(  \frac{2l_{a}}{a_{0}%
h_{rms}^{\prime}}\right)  ^{3/2}\label{PR_parameter}%
\end{equation}
and for example, for $l_{a}/a_{0}=0.05$, $h_{rms}^{\prime}=0.1$, our criterion
(\ref{new-criterion}) gives $\left(  h_{rms}/a_{0}\right)  _{th}=\frac
{4}{\sqrt{\pi}0.45}\frac{0.05}{0.1}=\allowbreak2.5$ while PR's one $\left(
h_{rms}/a_{0}\right)  _{th}=\sqrt{\alpha_{N}}$, and therefore they seem to
coincide for low bandwidth parameter, while at large bandwidths, it would be
important to further check results. It is remarkable that we obtained with a
very simple asperity model with exponential distribution of heights
(Ciavarella, 2017c)
\begin{equation}
\frac{h_{rms}}{a_{0}}<0.33\left(  \frac{l_{a}}{a_{0}h_{rms}^{\prime}}\right)
^{3/2}\label{Chi_parameter}%
\end{equation}
which is remarkably close both qualitatively and quantitatively to PR
parameter (\ref{PR_parameter}) at low bandwidths: therefore, while all
criteria seem to qualitatively give similar results in the limit case of low
bandwidth, the details differ at large bandwidth, and in this case, there
remains some uncertainty also because PR simulations show a threshold for
stickiness which is not corresponding to their own data on pull-off as
discussed in various previous papers (Ciavarella 2017a,b,c, Ciavarella \&
Papangelo 2017b, Ciavarella \& Papangelo 2017c).

\section{\bigskip Examples}

Let us consider a self-affine surface with power law PSD $A\left\vert
\mathbf{q}\right\vert ^{-2\left(  1+H\right)  }$ for wavevectors
$q_{0}<\left\vert \mathbf{q}\right\vert <q_{s}$ ($q=2\pi/\lambda$) and zero
otherwise (pure power-law). The surfaces have Hurst exponent $H=0.8$, and
$\lambda_{0}=2048a_{0}$, where $a_{0}$ could be an atomic spacing, but more in
general here enters only as a normalizing factor for the energy of adhesion
which we define as $l_{a}/a_{0}=0.05$ when we imitate the Lennard-Jones
potential (see also Pastewka-Robbins (2014)). For slopes we use $h_{rms}%
^{\prime}=0.1$ unless otherwise indicated (and notice that there is a minimum
level of slope for a given rms amplitude as we have fixed the smallest
wavevector). Using (\ref{JKR-final}), we obtain some example results.

\begin{center}%
\begin{tabular}
[c]{ll}%
{\includegraphics[
height=2.7132in,
width=4.8285in
]%
{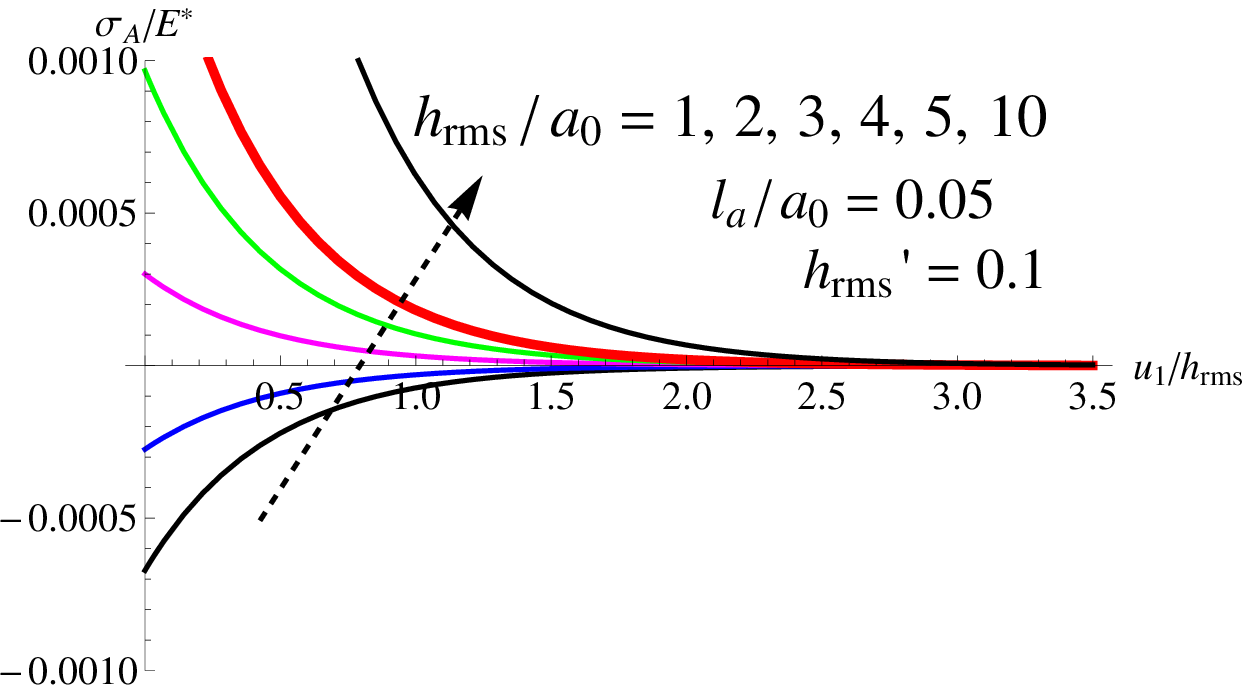}%
}
& (a)\\%
{\includegraphics[
height=2.9398in,
width=4.8285in
]%
{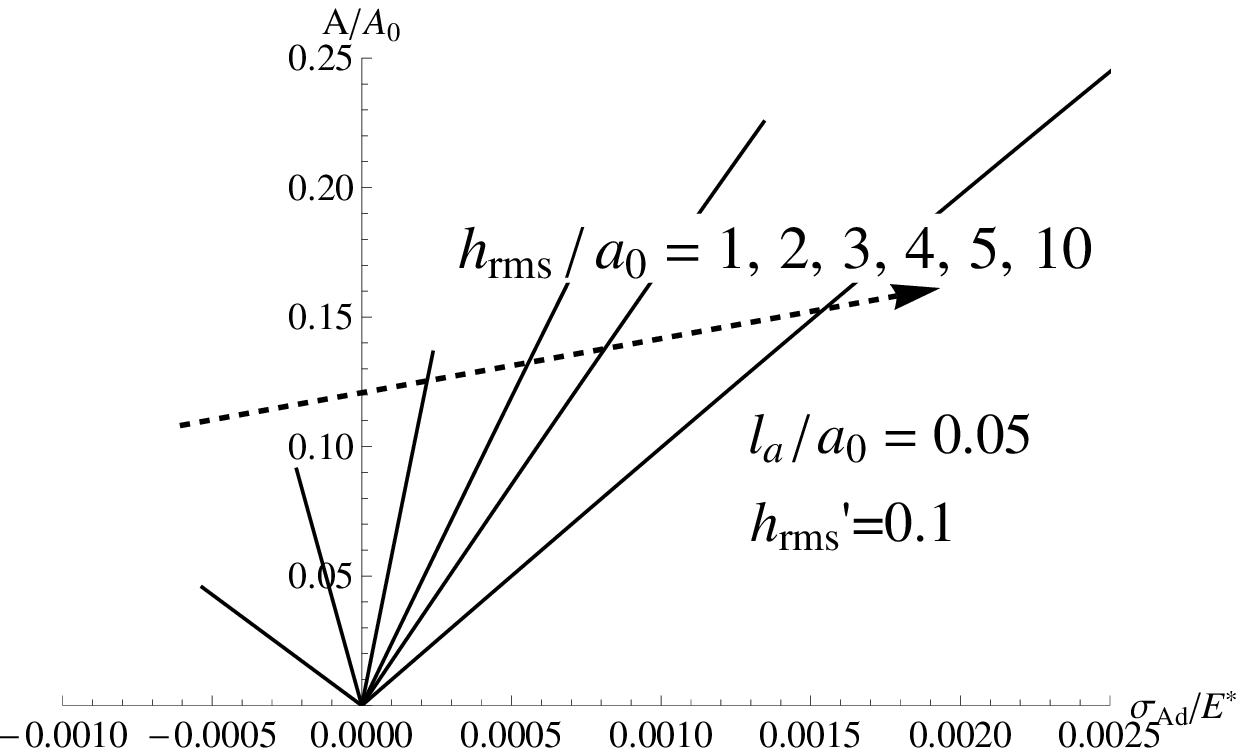}%
}
& (b)\\
&
\end{tabular}

\begin{tabular}
[c]{ll}%
{\includegraphics[
height=3.2179in,
width=5.056in
]%
{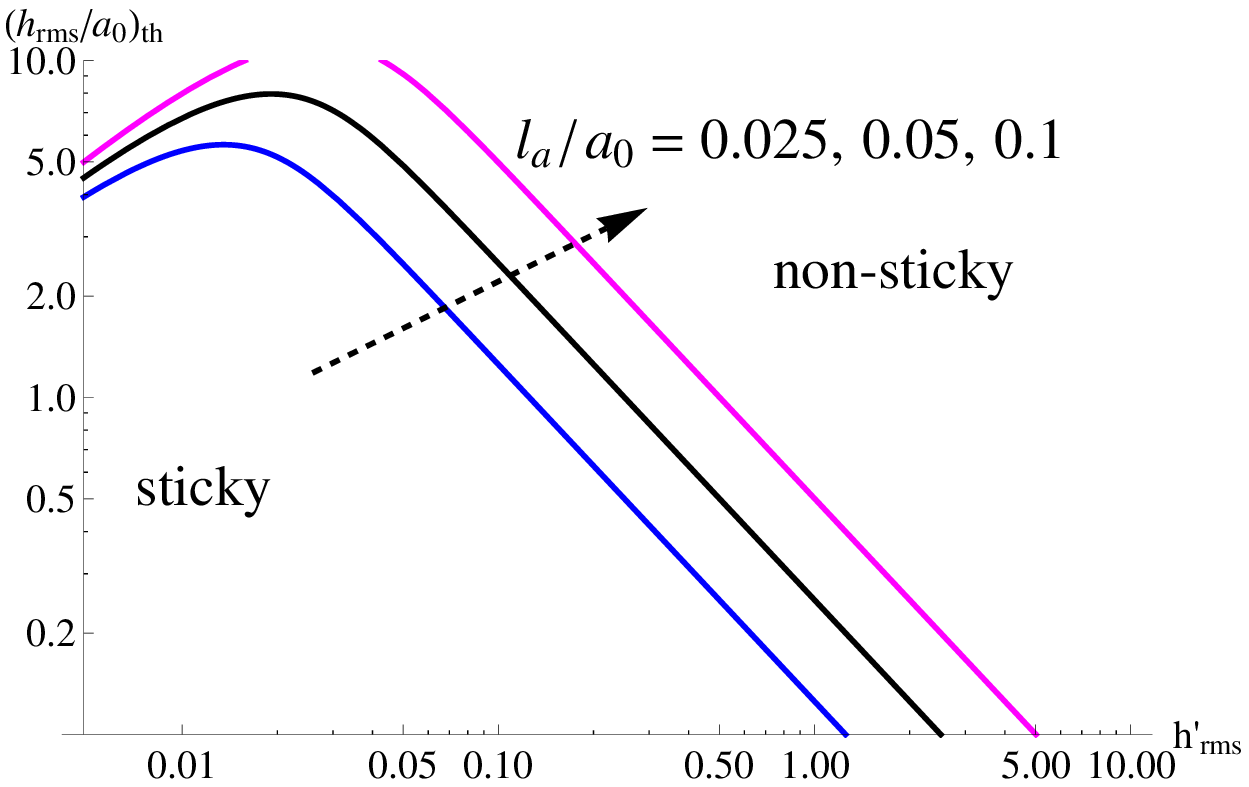}%
}
& (c)
\end{tabular}

Fig.2. (a) The adhesive mean stress vs the mean gap $u_{1}$ and (b) fraction
area over the nominal one $A/A_{0}:$ for various levels of $h_{rms}/a_{0}$ for
adhesion level $l_{a}/a_{0}=0.05$ and rms slopes $h_{rms}^{\prime}=0.1$. (c)
threshold "sticky"-"non-sticky" in terms of $h_{rms}/a_{0}$ as a function of
$h_{rms}^{\prime}$ and for various adhesion levels $l_{a}/a_{0}%
=0.025,0.05,0.1$. 
\end{center}

Fig.2a,b show that with increasing rms amplitude, the curves become
increasingly less adhesive as expected, for a given $h_{rms}^{\prime}=0.1$.
The curves in Fig.2b show a linear trend both in the "unsticky" and "sticky"
cases. Notice that the contact area, even at $u_{1}=0$ are still quite small.
Both these effects are due, again, to having used the asymptotic simple
Persson solution at small pressures.

Fig.2c shows that the "threshold" for stickiness $\left(  h_{rms}%
/a_{0}\right)  _{th}$ is, apart from an initial range in which we don't have a
reliable result since the slopes are too small to apply the Persson's result
for self-affine processes (we have too small bandwidth), generally there is a
very good power law regime 

Fig.3 show the decay of the crude estimates of pull-off (estimated as the
value at zero gap) with rms amplitude, for various $h_{rms}^{\prime
}=0.05,0.1,0.2$. This is qualitatively in agreement with previous results
(Ciavarella, 2017a). However, in the present case, there is a dependence also
on the slopes $h_{rms}^{\prime}$, whereas Ciavarella (2017a) only involved the
rms amplitudes $h_{rms}$, which seemed to fit better the case of the
Pastewka-Robbins (2014) simulations. However, this may simply mean that being
those simulations concerned with roughness at nanoscale with very low Tabor
parameter (of the order of 1), the DMT "bearing-area" model of Ciavarella
(2017a) which only involved the rms amplitudes $h_{rms}$, is a better model
for this case. Unfortunately, it is difficult to find in the literature
accurate solutions of the JKR problem with roughness, except for (Carbone et
al., 2015) which however, are rather limited to very few results.

\begin{center}%
\begin{tabular}
[c]{ll}%
{\includegraphics[
height=3.0079in,
width=5.0571in
]%
{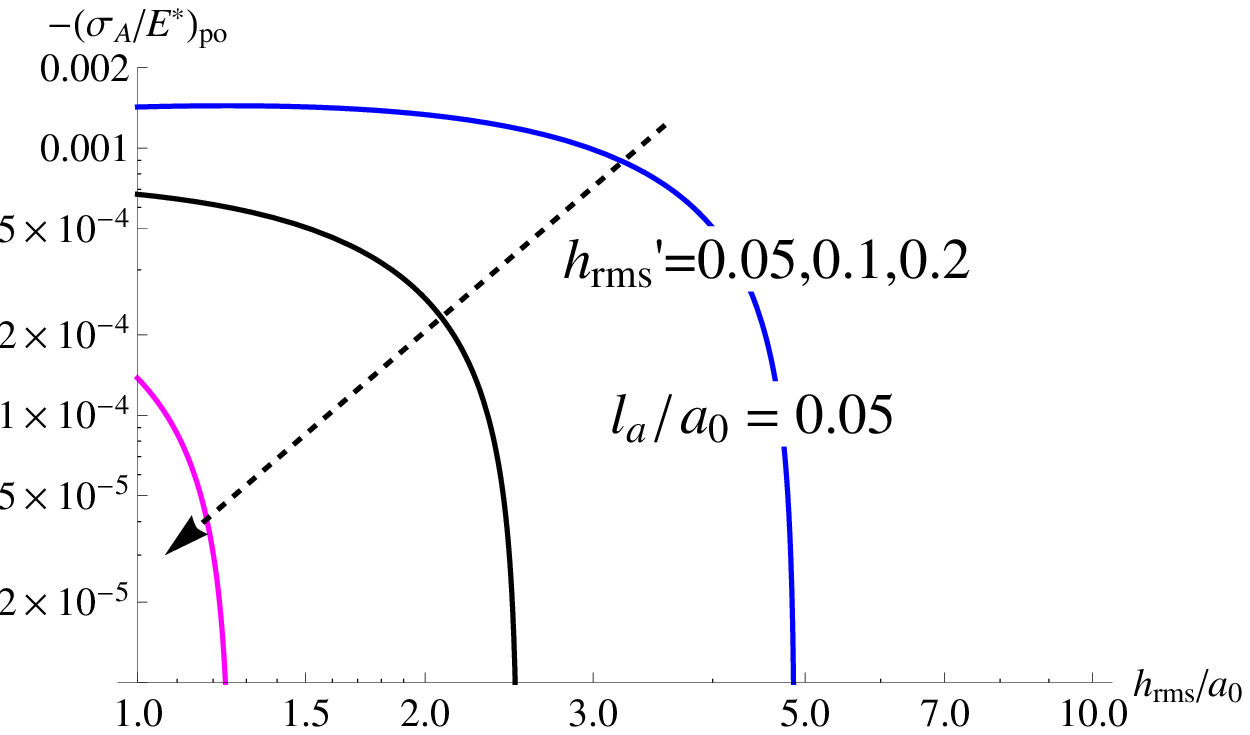}%
}
& (a)\\%
{\includegraphics[
height=3.1059in,
width=5.0571in
]%
{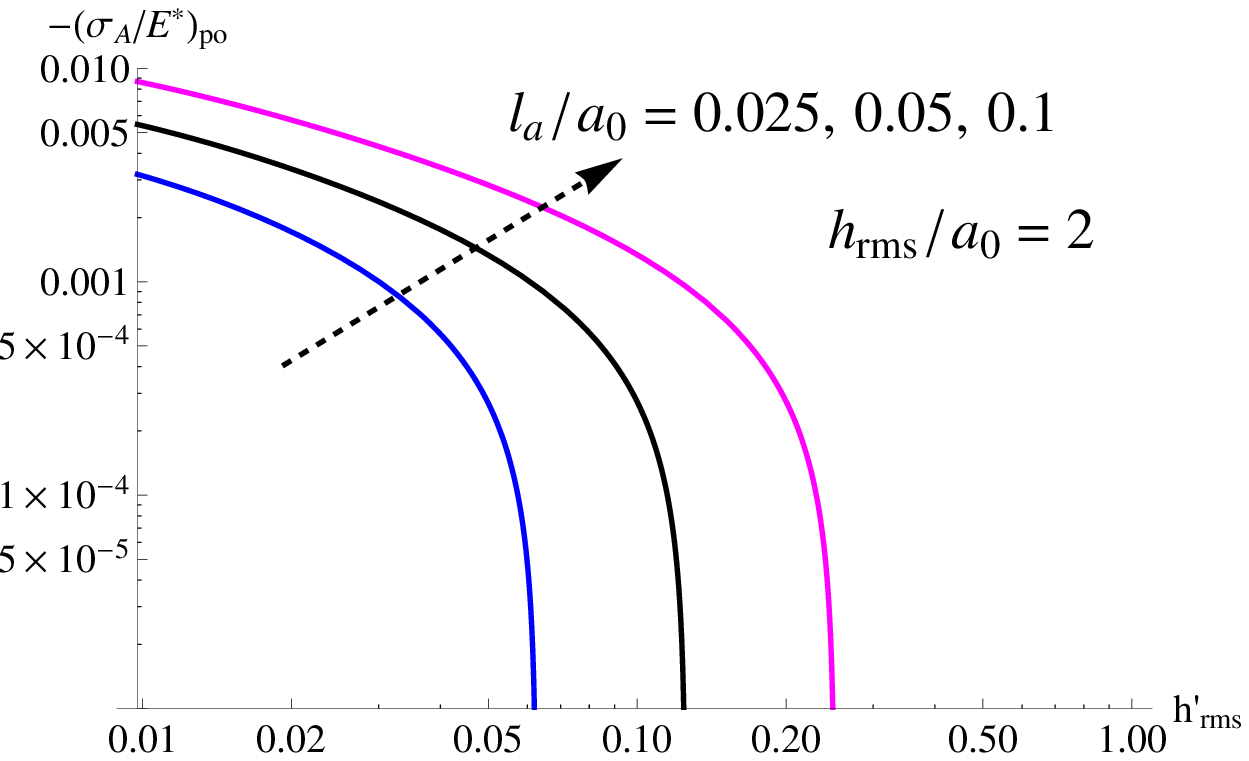}%
}
& (b)
\end{tabular}

Fig.3. A very crude estimate of pull-off value. (a) as a function of
$h_{rms}/a_{0}$ and $l_{a}/a_{0}=0.05$ while $h_{rms}^{\prime}=0.05,0.1,0.2$.
(b) for a given $h_{rms}/a_{0}=2$ and as a function of $h_{rms}^{\prime}$ for
various values of $l_{a}/a_{0}=0.025,0.05,0.1$
\end{center}

\section{\bigskip Conclusion}

We have provided a very simple model for JKR regime of a general contact,
within the approximation that contact areas with adhesion are the same as
contact area under adhesiveless conditions but with an appropriately increased
load. This has led to a very simple solution, which only needs the variation
of the contact area and load with indentation in the adhesiveless problem, for
which powerful analytical or numerical methods are available for many users,
whereas adhesive codes are much more complex, particular those involving
highly non-linear force laws like Lennard-Jones. The solution gives exact
results in axisymmetric cases, sinusoidal case, and similar results in an
ideal asperity model, resembling closely to the DMT model of Fuller and Tabor
which therefore is similar to the unloading prediction of a JKR asperity
model. It seems to give reasonable results even for the very complex problem
of JKR adhesion for rough surfaces, for which no analytical previous result is
known, at negative loads. However, it should be borne in mind that the JKR
limit is increasingly inappropriate at small scales. Near pull-off, we have
provided only the crudest estimates, since the more accurate values should
require the non-asymptotic form of the Persson's adhesiveless solution. The
quantitative comparison with an ideal test case using an asperity model (which
does not represent an actual random roughness as we know) shows that the
transition from sticky to non-sticky seems correctly predicted, although$\ $%
the effect of adhesion may be overestimated at low surface energy, while
underestimated at high ones. 

\section{\bigskip Acknowledgements}

The author is grateful to Prof.JR Barber (Michigan) for extensive discussions.
Also, Dr. A.\ Papangelo (TUHH, Hamburg) for initial suggestions regarding the
JKR methodology.

\section{\bigskip References}

Ciavarella, M., \& Papangelo, A. (2017). A modified form of Pastewka--Robbins
criterion for adhesion. The Journal of Adhesion, 1-11.

Ciavarella, M. (2015). Adhesive rough contacts near complete contact.
International Journal of Mechanical Sciences, 104, 104-111.

Afferrante, L., Ciavarella, M., \& Demelio, G. (2015). Adhesive contact of the
Weierstrass profile. In Proc. R. Soc. A (Vol. 471, No. 2182, p. 20150248). The
Royal Society.

Carbone, G., Pierro, E., \& Recchia, G. (2015). Loading-unloading hysteresis
loop of randomly rough adhesive contacts. Physical Review E, 92(6), 062404.

Ciavarella, M., Xu, Y., \& Jackson, R. L. (2018). Some Closed-Form Results for
Adhesive Rough Contacts Near Complete Contact on Loading and Unloading in the
Johnson, Kendall, and Roberts Regime. Journal of Tribology, 140(1), 011402.

Ciavarella, M., Greenwood, J. A., \& Barber, J. R. (2017). Effect of Tabor
parameter on hysteresis losses during adhesive contact. Journal of the
Mechanics and Physics of Solids, 98, 236-244.

Ciavarella, M. (2017a) A very simple estimate of adhesion of hard solids with
rough surfaces based on a bearing area model. Meccanica, 1-10.

Ciavarella, M. (2017b). On the use of DMT approximations in adhesive contacts,
with remarks on random rough contacts. Tribology International, 114, 445-449.

Ciavarella, M. (2017c). On Pastewka and Robbins' Criterion for Macroscopic
Adhesion of Rough Surfaces. Journal of Tribology, 139(3), 031404.

Ciavarella, M., Papangelo, A., \& Afferrante, L. (2017). Adhesion between
self-affine rough surfaces: Possible large effects in small deviations from
the nominally Gaussian case. Tribology International, 109, 435-440.

Ciavarella, M., \& Papangelo, A. (2017a). A random process asperity model for
adhesion between rough surfaces. Journal of Adhesion Science and Technology, 1-23.

Ciavarella, M., \& Papangelo, A. (2017b). A generalized Johnson parameter for
pull-off decay in the adhesion of rough surfaces. Phys Mesomech, 20(5).

Ciavarella, M., \& Papangelo, A. (2017c). A modified form of Pastewka--Robbins
criterion for adhesion. The Journal of Adhesion, 1-11.

Ciavarella, M., \& Papangelo, A. (2017d). On the sensitivity of adhesion
between rough surfaces to asperity height distribution. Phys Mesomech, 20(5).

\bigskip Derjaguin, B. V., Muller V. M. \& Toporov Y. P. (1975). Effect of
contact deformations on the adhesion of particles. J. Colloid Interface Sci.,
53, pp. 314--325

Fuller, K. N. G., \& Tabor, D. F. R. S. (1975). The effect of surface
roughness on the adhesion of elastic solids. In Proceedings of the Royal
Society of London A: Mathematical, Physical and Engineering Sciences (Vol.
345, No. 1642, pp. 327-342). The Royal Society.

Guduru, P. R. (2007). Detachment of a rigid solid from an elastic wavy
surface: theory. Journal of the Mechanics and Physics of Solids, 55(3), 445-472.

Johnson, K. L., \& Greenwood, J. A. (2005). An approximate JKR theory for
elliptical contacts. Journal of Physics D: Applied Physics, 38(7), 1042.

\bigskip Johnson, K. L., K. Kendall, and A. D. Roberts. (1971). Surface energy
and the contact of elastic solids. Proc Royal Soc London A: 324. 1558.

\bigskip Johnson, K. L. (1995). The adhesion of two elastic bodies with
slightly wavy surfaces. Int. J. Solids Structures, 32\textbf{\ (}No.
3/4\textbf{)}, 423-430.

Papangelo, A., Hoffmann, N., \& Ciavarella, M. (2017). Load-separation curves
for the contact of self-affine rough surfaces. Scientific reports, 7(1), 6900.

Pastewka, L., \& Robbins, M. O. (2014). Contact between rough surfaces and a
criterion for macroscopic adhesion. Proceedings of the National Academy of
Sciences, 111(9), 3298-3303.

Persson, B. N. (2001). Theory of rubber friction and contact mechanics. The
Journal of Chemical Physics, 115(8), 3840-3861.

Persson, B.N.J., Albohr, O., Tartaglino, U., Volokitin, A.I., Tosatti, E.,
(2005). On the nature of surface roughness with application to contact
mechanics, sealing, rubber friction and adhesion. J. Phys.: Condens. Matter.
17, 1--62.

Persson, B. N. J. "Adhesion between an elastic body and a randomly rough hard
surface." (2002) The European Physical Journal E: Soft Matter and Biological
Physics 8, no. 4 : 385-401.

Persson, B. N. J., \& Tosatti, E. (2001). The effect of surface roughness on
the adhesion of elastic solids. The Journal of Chemical Physics, 115(12), 5597-5610.

Persson, B. N., \& Scaraggi, M. (2014). Theory of adhesion: Role of surface
roughness. The Journal of chemical physics, 141(12), 124701.

Popov, V. L., \& He\ss , M. (2015). Method of dimensionality reduction in
contact mechanics and friction. Springer Berlin Heidelberg.

Putignano, C., Afferrante, L., Carbone, G., \& Demelio, G. (2012). A new
efficient numerical method for contact mechanics of rough surfaces.
International Journal of Solids and Structures, 49(2), 338-343.

\end{document}